\documentclass[aps,pra,numerical,superscriptaddress,showpacs,
reprint
]{revtex4-1}
\usepackage{amsmath}
\usepackage{color}
\usepackage{graphicx}
\usepackage[english]{babel}
\usepackage[noload=abbr]{siunitx}
\usepackage{subcaption}

\begin{document}
\title{Fiber-integrated spectroscopy device for hot alkali vapor}

\author{Josephine Gutekunst}
\affiliation{5.~Physikalisches Institut and Center for Integrated Quantum Science and Technology, University of Stuttgart, Pfaffenwaldring 57, 70569 Stuttgart, Germany}
\author{Daniel Weller}
\affiliation{5.~Physikalisches Institut and Center for Integrated Quantum Science and Technology, University of Stuttgart, Pfaffenwaldring 57, 70569 Stuttgart, Germany}
\author{Harald K\"ubler}
\affiliation{5.~Physikalisches Institut and Center for Integrated Quantum Science and Technology, University of Stuttgart, Pfaffenwaldring 57, 70569 Stuttgart, Germany}
\author{Jan-Philipp Negel}
\affiliation{Institut f\"ur Strahlwerkzeuge (IFSW), University of Stuttgart, Pfaffenwaldring 43, Stuttgart, Germany}
\author{Marwan Abdou Ahmed}
\affiliation{Institut f\"ur Strahlwerkzeuge (IFSW), University of Stuttgart, Pfaffenwaldring 43, Stuttgart, Germany}
\author{Thomas Graf}
\affiliation{Institut f\"ur Strahlwerkzeuge (IFSW), University of Stuttgart, Pfaffenwaldring 43, Stuttgart, Germany}
\author{Robert L\"ow}
\email{r.loew@physik.uni-stuttgart.de}
\affiliation{5.~Physikalisches Institut and Center for Integrated Quantum Science and Technology, University of Stuttgart, Pfaffenwaldring 57, 70569 Stuttgart, Germany}

\date{\today} 

\begin{abstract}
We introduce an all-glass, vacuum tight, fiber-integrated and alkali compatible spectroscopy device consisting of two conventional optical fibers spliced to each end of a capillary. This is mainly realized through a decentered splicing method allowing to refill the capillary and control the vapor density inside. We analyze the light guidance of the setup through simulations and measurements of the transmission efficiency at different wavelengths and show that filling it with highly reactive alkali metals is possible and that the vapor density can be controlled reliably.
\end{abstract}

\pacs{}

\maketitle

\section{Introduction}
Many laser based applications,
for example high-precision frequency measurements,
electromagnetic field sensing and non-linear and quantum optics use gas-phase materials.
To implement these in practical and compact devices hollow-core fibers show a promising approach.
They can be filled with various gases and have a large overlap between the light field and the atomic ensemble. 
Gas-based all-fiber devices consisting of a hollow-core photonic crystal fiber (HC-PCF) 
with a conventional optical single-mode fiber spliced at each end have been reported in \cite{Russel}.
Here the fibers were loaded with the active gas through the use of helium as a buffer gas to reduce contamination during the splicing process. They could be successfully filled with acetylene or hydrogen, but loading them with alkali metals was challenging due to the high reactivity of the alkali. That would however be of great interest since alkali atoms are spectrally indistinguishable and therefore very suitable for reference measurements, non-linear optics like the storage of light \cite{Lukin} and for sensing applications \cite{Budker.2007}.
For example the base unit of time, the second, is defined through a transition frequency in cesium. 
Nevertheless, there have already been different approaches to fill fibers with alkali metals. 
In Ref. \cite{Hofferberth} the hollow-core fibers were filled with laser-cooled rubidium atoms to achieve non-linear optics at a very low light level and in Ref. \cite{Blatt} a high optical density for ultra-cold atoms inside a fiber has been achieved. However, the filling process with ultra-cold atoms is demanding and it is not feasible to achieve miniaturized and integrable tools. Abandoning the cooling and loading the fiber with room-temperature atoms is a promising step in this direction and has already been demonstrated with rather bulky alkali reservoirs including vacuum chambers made of steel. \cite{Benabid2,Wamsley2,Gaeta2}. Aside from that it is sometimes also advantageous to have all-glass vapor cells operating at room-temperature, for example if high densities are required as it is the case for applications like electromagnetic field sensing \cite{Shaffer}. 
In this paper we introduce an all-glass, fiber-integrated device, which can be filled with a hot alkali vapor by means of a newly developed decentered splicing method. This method leaves a gap through which the atoms can pass to fill and refill the fiber from an attached reservoir. In the following the manufacturing process is described in detail and the proof of principle is produced through filling the device successfully with rubidium and performing absorption spectroscopy with it.

\section{Manufacturing Process}
\begin{figure}[ptbh]
	\centering
	\includegraphics[width=\linewidth]{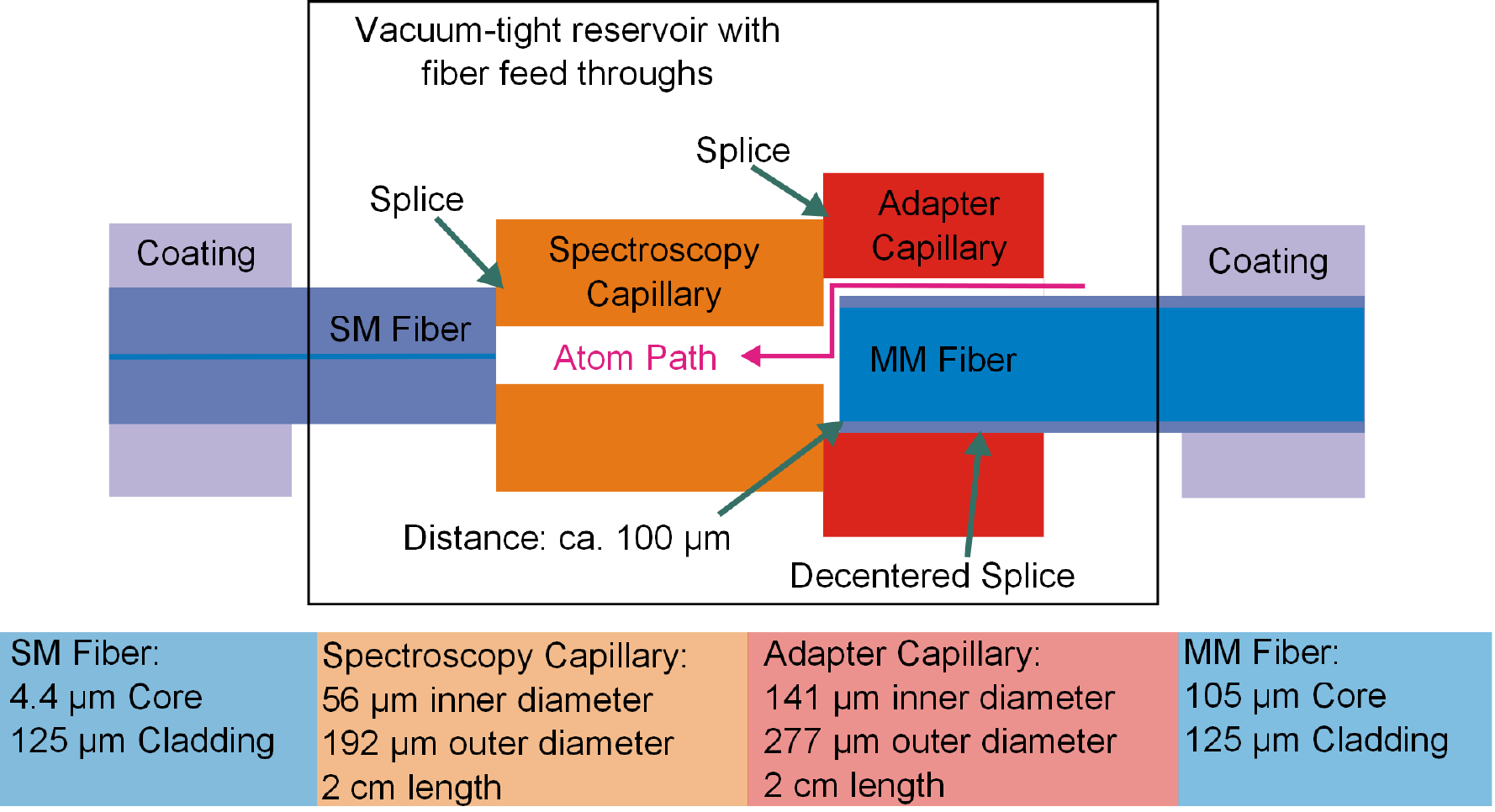}
	\caption{Schematic overview (not to scale) of the fiber-integrated device for the spectroscopy measurements. The light is injected from the left hand side through a single-mode fiber (SM Fiber). It propagates inside the spectroscopy capillary and after leaving the capillary is collected by a multi-mode fiber (MM Fiber). To allow the desired atoms to enter the spectroscopy capillary the MM fiber was spliced decentered into an adapter capillary and placed inside a reservoir.}
	\label{fig:SpliceSchematic}
\end{figure}
The basic concept is similar to the one shown in \cite{Russel}: a hollow-core fiber, in our case a capillary with a core diameter of 56 $\mu m$, is spliced to a conventional optical fiber at each end (see Fig.\ref{fig:SpliceSchematic}). The fiber array was designed at the base of the following criteria: 
\\
First of all, the probe beam has to be guided through the capillary with a large interaction cross-section between light and atoms (i.e. a large mode diameter in the capillary) and with an acceptable amount of losses to allow for spectroscopy. This can be achieved with splicing a single-mode (SM) fiber  to the capillary while optimizing the mode matching between the two.
\\ 
Second, the light has to be collected behind the capillary. Since the cross section of the modes in the capillary is large, a fiber with a wide core to collect all the light is desirable, therefore a multi-mode (MM) fiber is used.
\\
Third, the capillary needs to be filled with a controlled atmosphere of atoms. This is the most difficult part to achieve with alkali atoms because of their high reactivity. In previous attempts the fibers were filled under vacuum conditions and then the ends were sealed with a CO$_2$-laser. This did not lead to the desired outcome because of chemical reactions of the alkali with atoms inside the fiber walls during the sealing process and the density changing therefore non-reproducibly.
Another problem is that over longer time scales the alkali atoms diffuse into the glass walls, leading to a density loss. If this loss is countered with a very high initial density, the atoms build a thin film on the capillary walls, which reduces the light propagation. The solution to this problem is to have a reservoir from which the capillary can be refilled and thus the density controlled. This can be attained through a decentered splicing method, which we describe in the following.
\\
\begin{figure}[t]
\centering
\includegraphics[width=\linewidth]{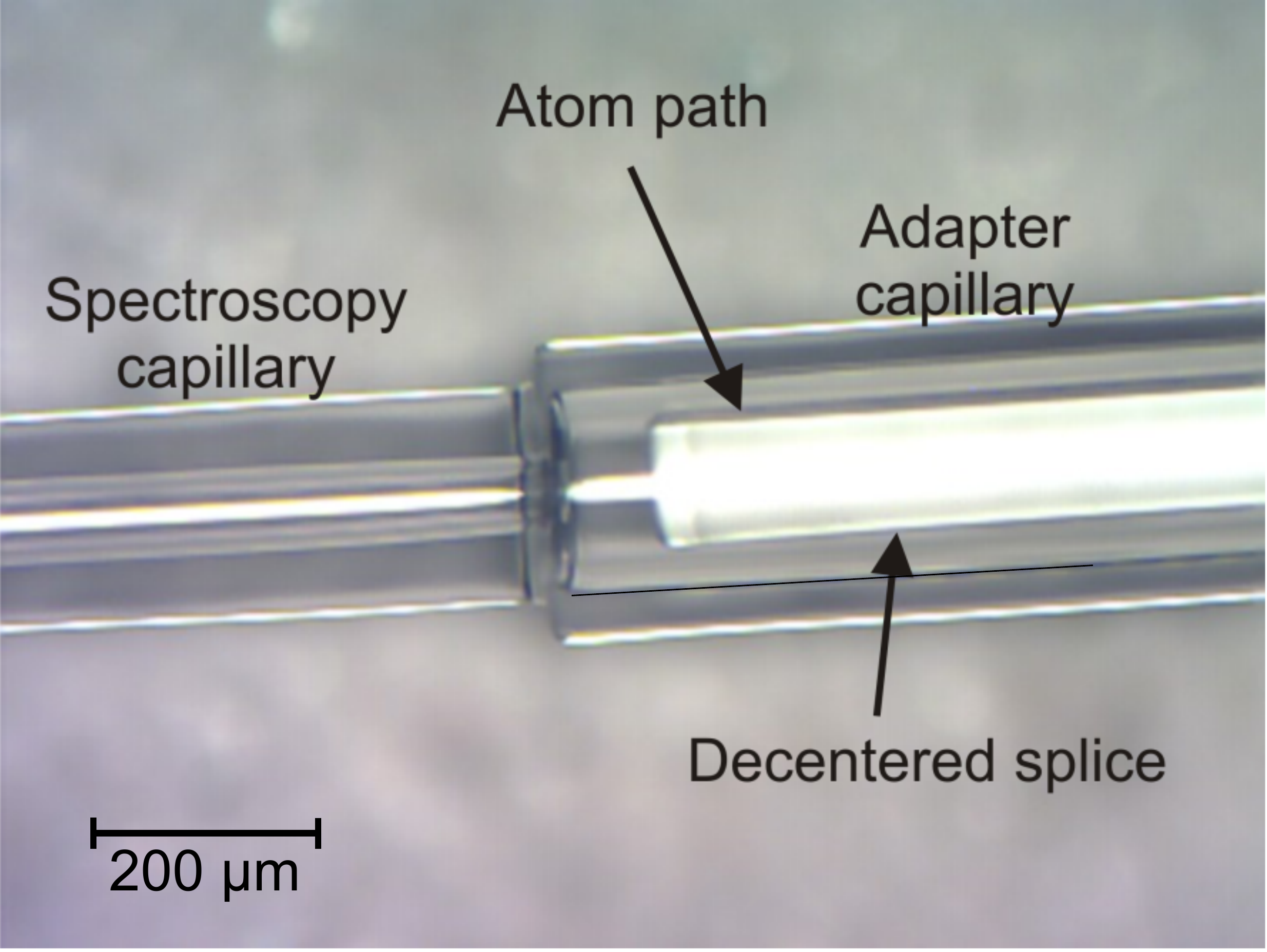}
\caption{Picture of the connection between MM fiber and spectroscopy capillary. The MM fiber was spliced on the adapter capillary in a decentered way leaving a path for the atoms to fill the spectroscopy capillary.}
\label{fig:SplicePicture}
\end{figure}
Fig.\ref{fig:SpliceSchematic} depicts a schematic overview of the used fiber assembly.
The light was injected into the spectroscopy capillary through a Thorlabs 780 HP single-mode fiber. It was made of fused silica and had an outer diameter of 125 $\mu$m and a core diameter of 4.4 $\mu$m. The outer diameter of the fused silica spectroscopy capillary was 192 $\mu$m, the inner diameter 56 $\mu$m and its length amounted to 2 cm. The SM fiber was directly spliced onto the spectroscopy capillary using a Fitel S184-PM fusion splicing device. Here, a mechanically stable and vacuum tight connection could be accomplished. Furthermore, the splicing parameters were adjusted in such a way that the inner part of the spectroscopy capillary did not collapse (which would eventually prevent guiding of light inside the capillary).
\\
At the end of the spectroscopy capillary the light was collected by a MM fiber. The fiber had a cladding diameter of 125 $\mu$m and a core diameter of 105 $\mu$m being substantially larger than the inner diameter of the spectroscopy capillary. The MM fiber was placed inside a 2 cm long adapter capillary with an outer diameter of 277 $\mu$m and an inner diameter of 141 $\mu$m (see Fig.\ref{fig:SplicePicture}). 
This difference of 16 $\mu$m between the cladding diameter of the MM fiber and the inner diameter of the adapter capillary allowed for using a decentered splicing method which opened a path for the atoms through the resulting channel at the side of the MM fiber. To achieve this, first the two capillaries were spliced on each other. Afterwards, the MM fiber was pushed into the adapter capillary with a remaining gap to the spectroscopy capillary of about 100 $\mu$m. The MM fiber was placed at the bottom of the adapter capillary and spliced in this position a few millimeters behind the connection of the two capillaries. This leaves an open path for the atoms to travel through (see Fig.\ref{fig:SpliceSchematic}). The fiber assembly was then placed inside a vacuum-tight reservoir. The fiberfeedthroughs were created relying on the different melting points for different glass materials, as described in detail in Ref. \cite{Weller:17}. 
\begin{figure}[bhtp]
	\centering
	\includegraphics[width=\linewidth]{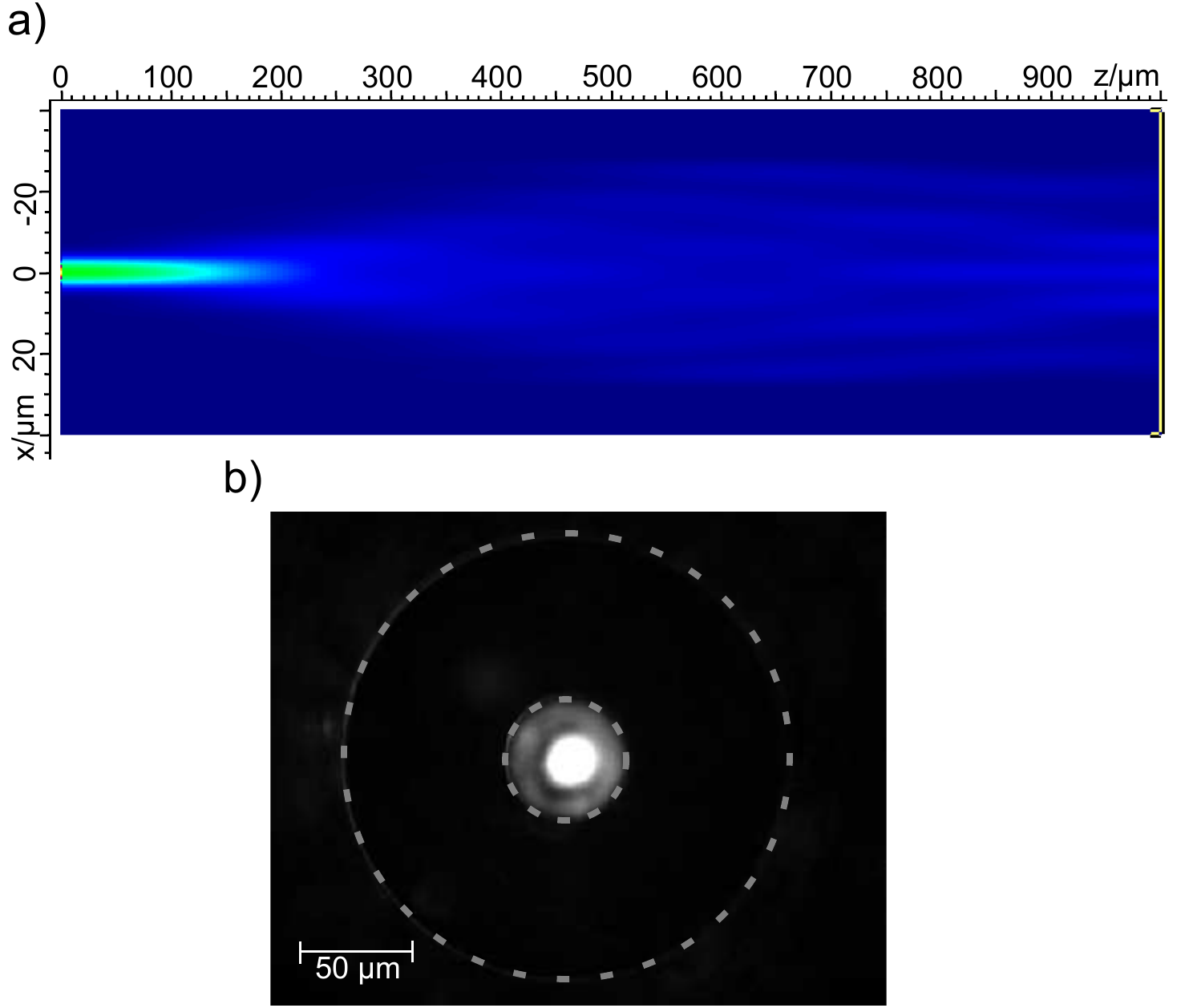}
	\caption{a) Simulated intensity distribution inside the spectroscopy capillary over a length of 1 mm after insertion through the SM fiber. b) Picture of the light emitted at the end facet of the spectroscopy capillary.}
	\label{fig:Simulation}
\end{figure}
\section{Light guidance}
The coupling of the light from the SM fiber to the spectroscopy capillary was simulated for light with a wavelength of 850 nm using the Software Fimmwave. 
The coupling of the light into the capillary is coherent and the light field inside the capillary can be represented by a linear combination of the fundamental and higher-oder modes in the basis of the capillary mode. Here, in a first step the modes inside the SM fiber and the capillary and in a second step the overlap integral between these modes were calculated. The calculated coupling efficiency from the fundamental mode of the SM fiber to the fundamental mode of the capillary was about $10\%$. 
The residual light was coupled to the higher-order part of the mode propagating inside the capillary, which leads to a coupling efficiency for all considered modes of about $60 \%$.
The resulting simulated intensity distribution over the first 1 mm of the spectroscopy capillary is depicted in Fig.\ref{fig:Simulation}a). 
It shows a combination of the fundamental mode and higher-order modes of the capillary, their field distributions reach further outward and show therefore a higher loss than the fundamental mode. However, a good confinement in the center of the capillary and a good interaction cross section was observed.
\\
The measured transmission efficiency of the light with a wavelength of 850 nm inside the capillary over a length of 2 cm was about $35\%$. This value originates most likely from hard to quantify higher-order losses inside the capillary in addition to the simulated coupling efficiency of the modes of $60~\%$. The measured intensity distribution at the end of the spectroscopy capillary can be seen in Fig.\ref{fig:Simulation}b).
\\
\begin{figure}[bh]
	\centering
	\includegraphics[width=\linewidth]{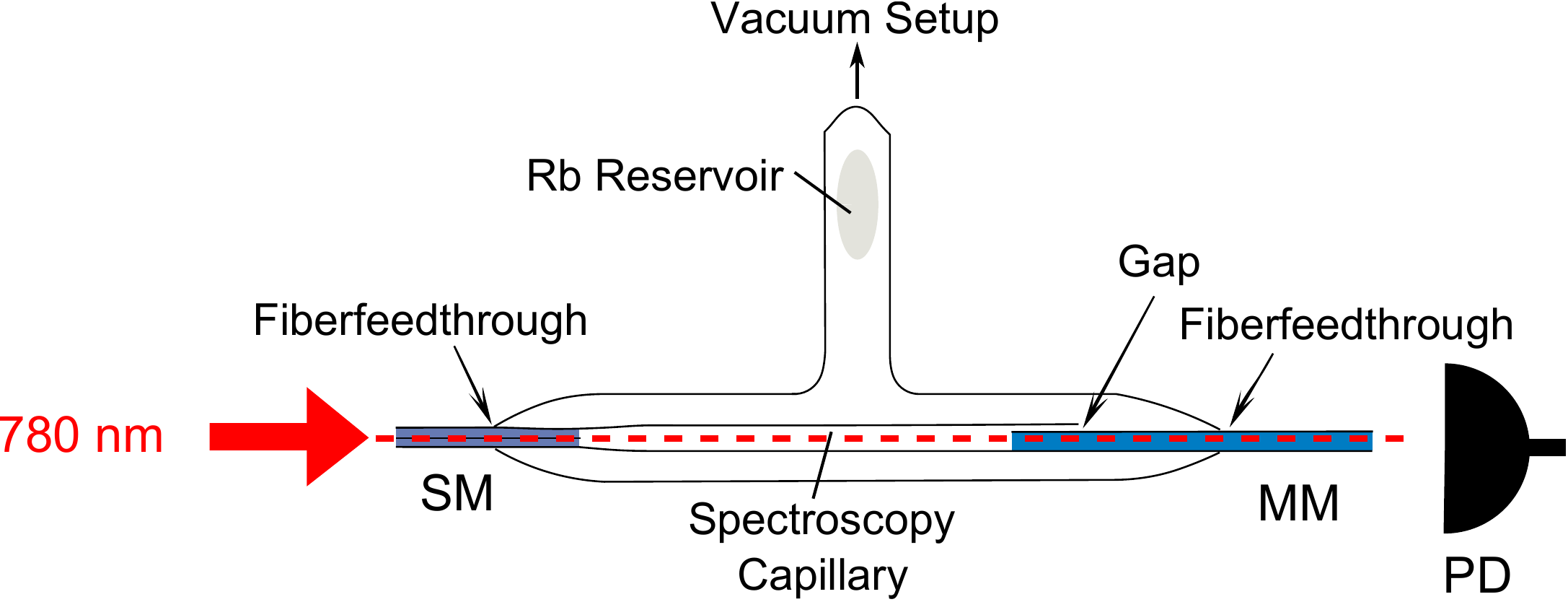}
	\caption{Device filled with rubidium. The rubidium diffused into the spectroscopy capillary through the gap resulting from the decentered splicing of the MM. The spectroscopy was done with 780 nm light coupled into the SM fiber and read out after the MM fiber with a photo-diode (PD).}
	\label{fig:SpectroscopySetup}
\end{figure}
The device was especially designed to work with the highly reactive alkali metals (which render the loading process challenging) but it can be used to do spectroscopy with various gases and liquids. In this case it was exemplary filled with rubidium (see section \ref{cap:spectroscopy}). 
The device is also versatile in respect to the used wavelength and light with a wavelength of 780 nm is used to drive the D2-line of rubidium. 
The measured transmission efficiency of the whole device (meaning the ratio of the laser power before coupling into the SM fiber and after coupling out of the MM fiber) for light with this wavelength was $5 \%$, being still enough for the spectroscopy purpose.
Coupling the light into the other direction lead to efficiencies two orders smaller of around $0.01 \%$. This is primarily due to the fact that in this direction the coupling efficiency into the capillary is worse and the SM fiber does not collect as much light from the capillary as the MM fiber does. The coupling efficiencies after the filling process stayed roughly the same.
\section{Spectroscopy}\label{cap:spectroscopy}
\begin{figure}[tb]
	\centering
	\includegraphics[width=\linewidth]{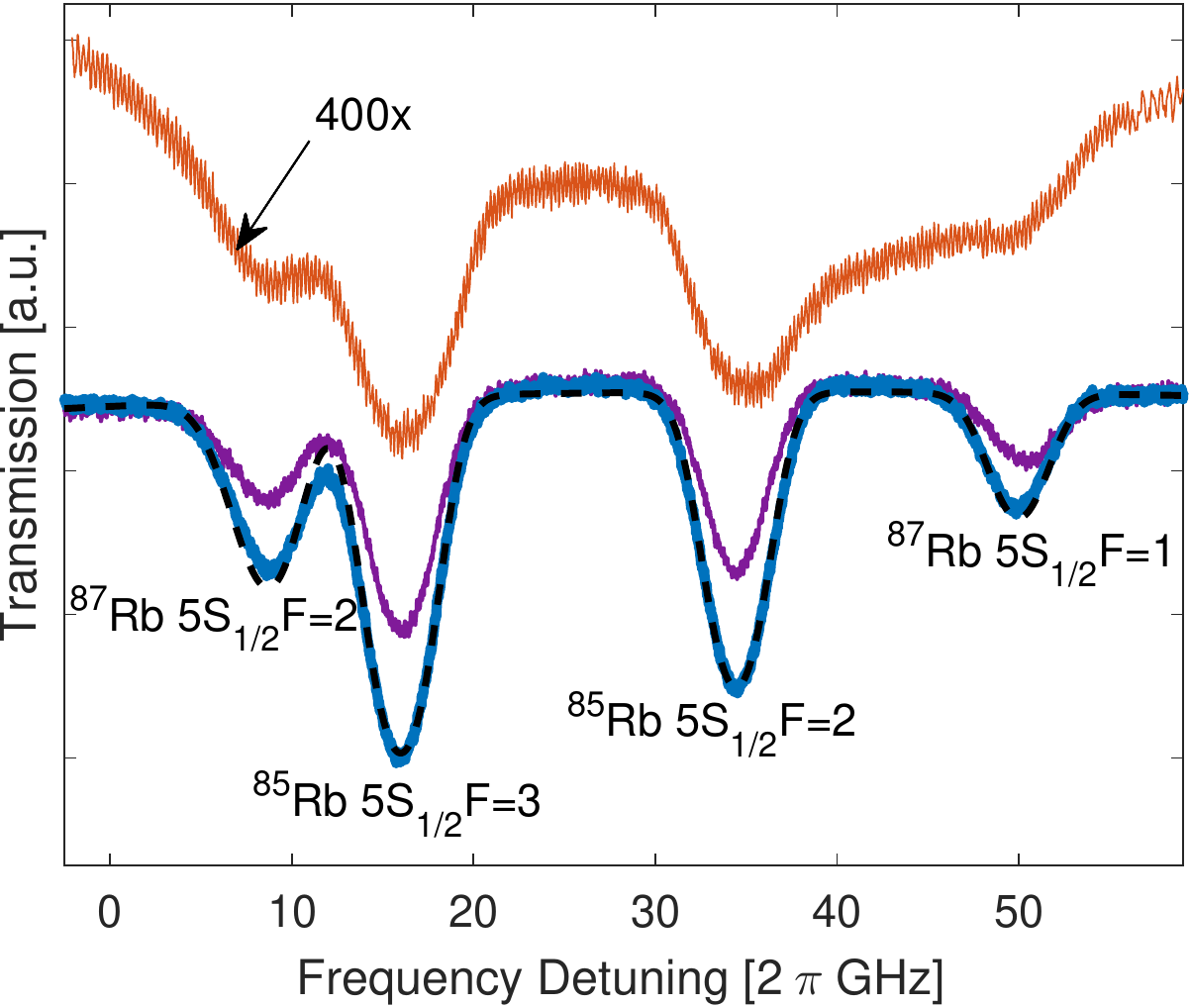}
	\caption{Absorption spectra inside the capillary for different coupling directions and reservoir temperatures $T_{\mathrm{res}}$. The ground state for each transition is marked inside the picture. The hyperfine splitting of the excited states can not be resolved due to Doppler broadening. Blue: Coupled into the SM fiber and read out after the MM fiber at $T_{\mathrm{res}} = 70~^\circ \mathrm{C}$. Violet: Coupled in the same direction at $T_{\mathrm{res}} = 45~^\circ \mathrm{C}$. Red: Coupled into the MM fiber and read out after the SM fiber at $T_{\mathrm{res}} = 70~^\circ \mathrm{C}$. The signal has been upscaled to fit the same transmission region as the other signal for better comparison. Dashed black: Gaussian fit summed over all hyperfine transitions.
		}
	\label{fig:Absorption}
\end{figure}
\begin{figure}[tbp]
	\centering
	\includegraphics[width=\linewidth]{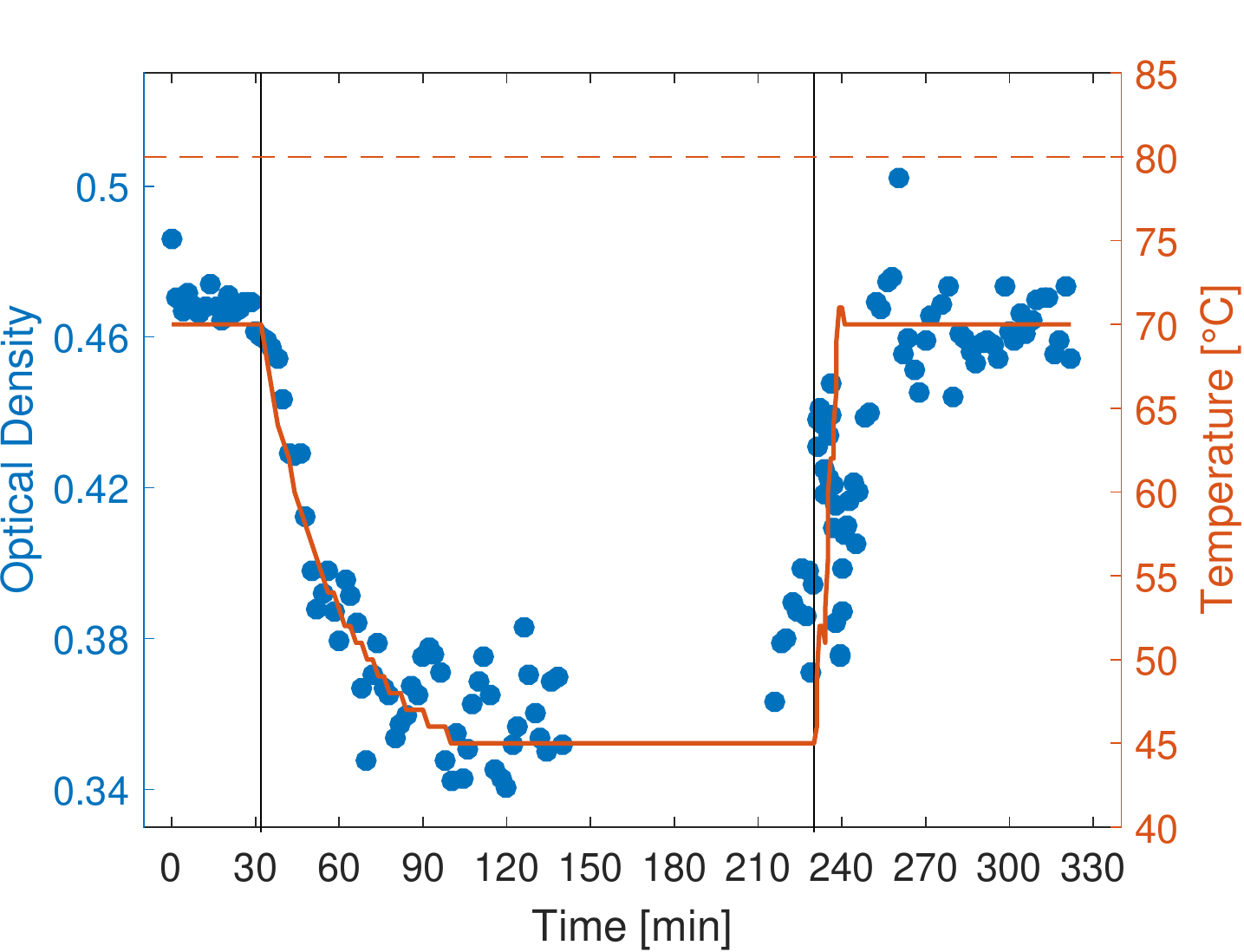}
	\caption{Adjustment of the optical density through temperature control. Blue dots: Optical density inside the capillary. Solid red line: Temperature inside the reservoir. Dashed red line: Temperature inside the capillary.}
	\label{fig:OD}
\end{figure}
To fill the capillary with rubidium, the capillary was first attached to a vacuum setup via the tube attached orthogonal to the fiber-array seen in Fig.\ref{fig:SpectroscopySetup} where also a ampule with liquid rubidium was attached to. It was then evacuated to a high vacuum of approximately $10^{-7} \mathrm{mbar}$ before opening the ampule and guiding a  liquid rubidium droplet into the handle and detaching it from the vacuum setup. 
This droplet can now be used as a reservoir to control the density inside the capillary and refill it through the gap on the MM side.
\\
Afterwards the whole device was placed into an oven, which allowed to control the temperature independently for the part containing the capillary, and the reservoir. With two Pt100-elements and PID-controllers the parts of the device were kept at constant temperatures with different set-points:
The vapor pressure of the rubidium atmosphere is controlled by the reservoir temperature, which was set to 70 $^\circ$C. To avoid condensation of the alkali inside the capillary, the temperature of the active part of the device was kept slightly higher, at 80 $^\circ$C.
The curing of the fiber only takes a few days in comparison to several month reported in \cite{Gaeta,Wamsley} since it is miniaturized and therefore more easily heatable and cold spots can be avoided.
\\
Light with a wavelength of 780 nm was coupled into the fiber to drive the D2-line of the rubidium for the measurements and read out after the MM fiber using a photo-diode (PD). 
In Fig.\ref{fig:Absorption} the resulting absorption spectra can be seen for the two different coupling directions. In both directions a good signal can be achieved. The device is of course designed to be operated SM to MM direction but being able to work in both directions could allow for spectroscopy with counter-propagating beams like for example saturation spectroscopy. 
\\
The taken spectrum can be fitted with a simple Gaussian function, which is summed over all hyperfine transitions.
From this fit (see Fig.\ref{fig:Absorption}) the optical density inside the capillary can be derived. 
It can be controlled through the temperature control of the oven, which can be seen in Fig.\ref{fig:OD} where the temperature of the capillary is left constant while the temperature of the reservoir is scanned from $70 ^\circ\mathrm{C}$ to $45 ^\circ\mathrm{C}$ and back to $70 ^\circ\mathrm{C}$ again. When the temperature in the reservoir drops the rubidium vapor density inside the reservoir goes down as well. This leads to a diffusion from the capillary through the gap into the reservoir and therefore the density inside the capillary drops as well. In Fig.\ref{fig:OD} this behavior can be seen as the density follows the temperature pretty instantaneously. In the end the same optical density is reached as at the beginning. This shows that the density inside the fiber can be quickly and reliably adjusted through the temperature control. In future this could also be done by light induced atom desorption \cite{Gaeta,Wamsley}.
\begin{figure}[t]
	\begin{subfigure}{0.99 \linewidth}
		\includegraphics[width=\linewidth]{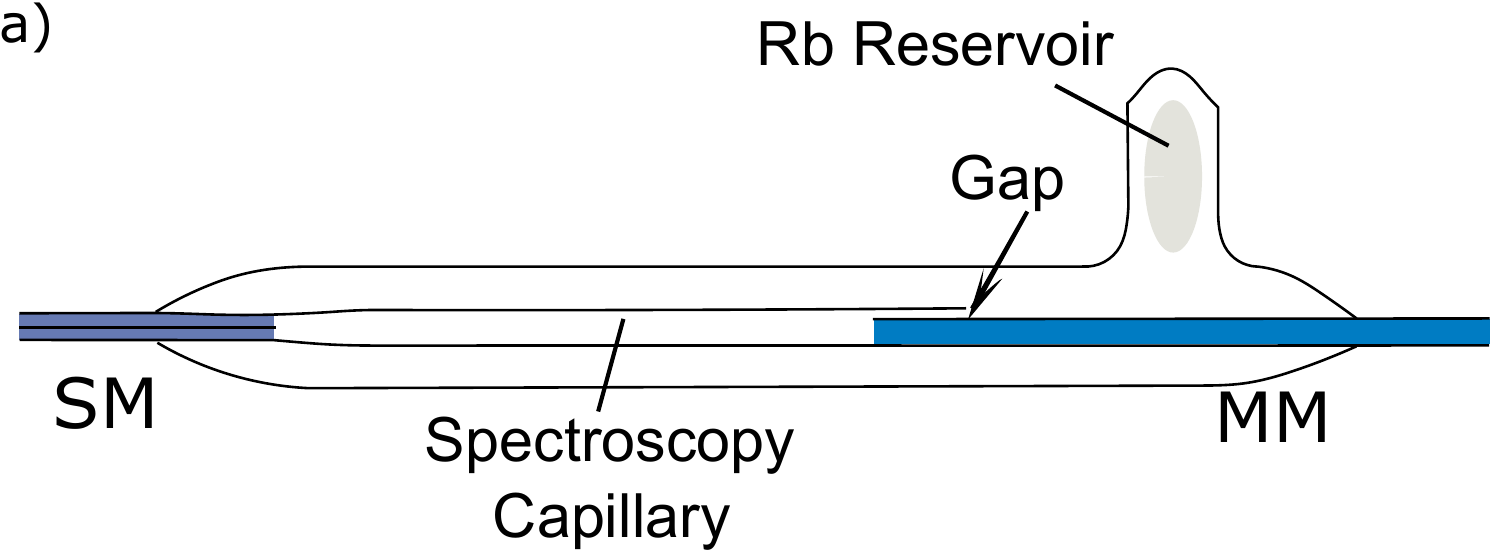}
	\end{subfigure}
	\begin{subfigure}{0.99 \linewidth}
		\includegraphics[width=\linewidth]{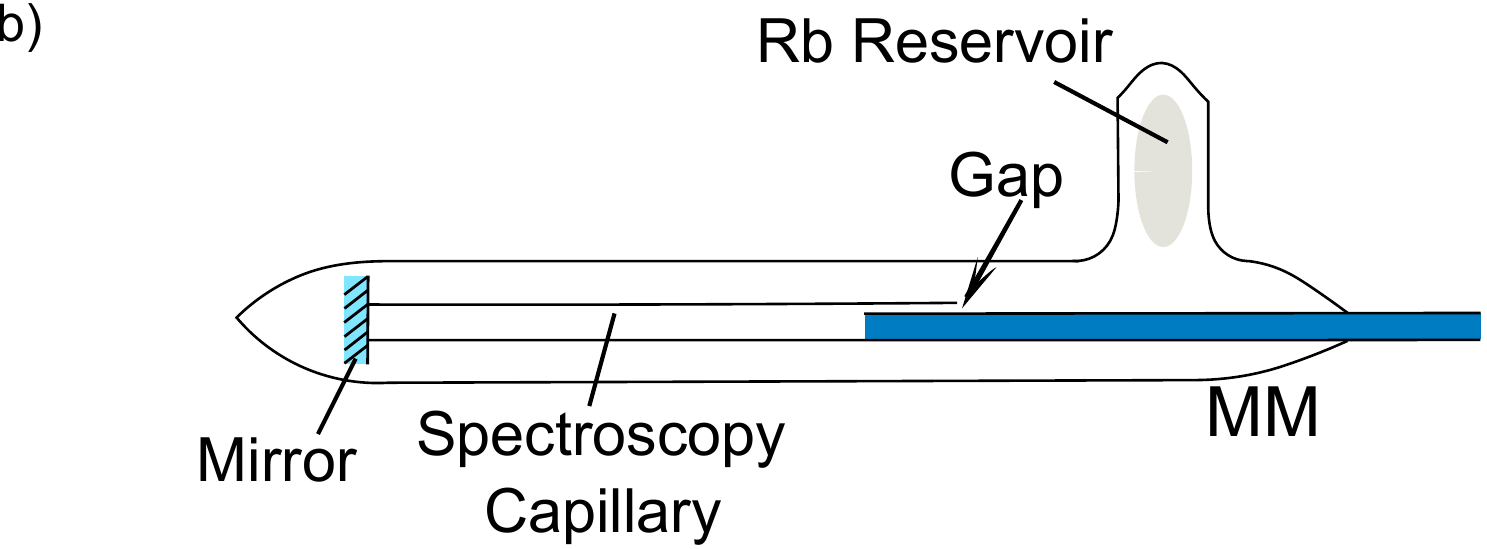}
	\end{subfigure}
	\begin{subfigure}{0.99 \linewidth}
		\includegraphics[width=\linewidth]{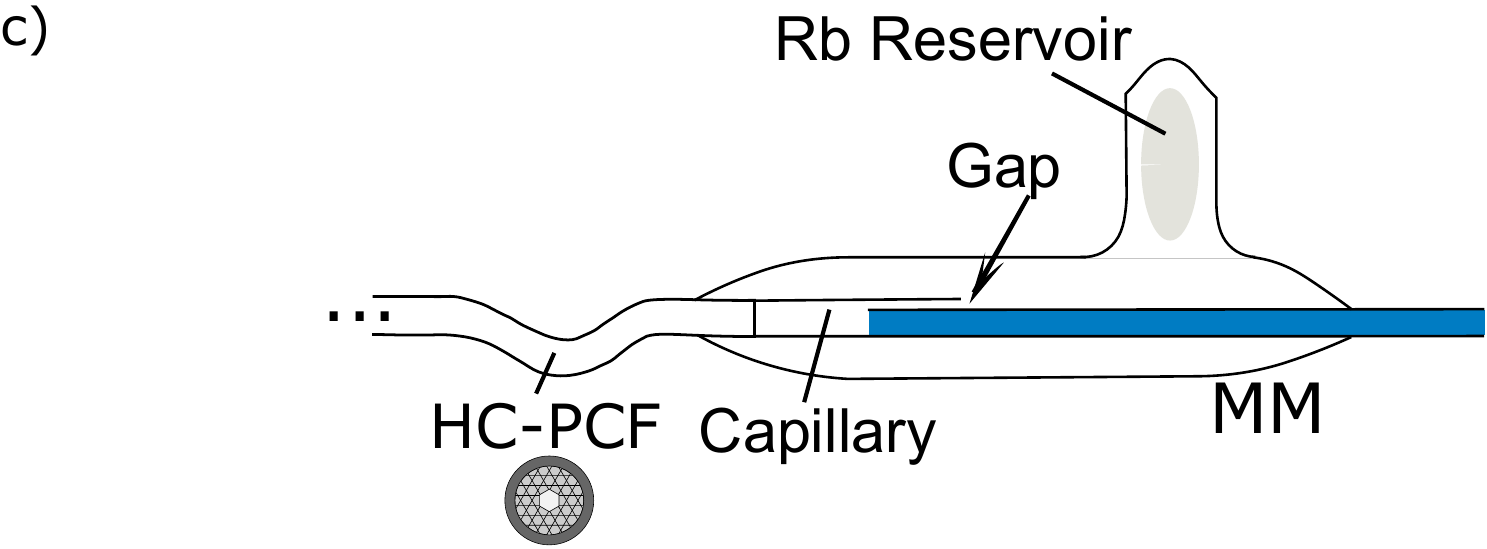}
	\end{subfigure}
	\caption{Three different possible further developments for the fiber-integrated device. }
	\label{fig:Outlook}
\end{figure}
\section{Outlook}
With the device described in this paper we have provided a proof of concept for an all-glass fiber-integrated spectroscopy device with hot alkali vapor. A sufficient lifetime of this type of device is expected, since we did not observe any degradation of its performance during an observation period of six months. 
We can now think about further improvements to get integrable devices. First of all it would be reasonable to get an obstruction-free spectroscopy area. This is easily attainable by moving the attachment place of the reservoir like it is for example shown in Fig.\ref{fig:Outlook}a). This could of course lead to different diffusion path lengths and therefore a change in optical density adjustment times but should show no difference in the other properties discussed.
\\
To achieve an antenna-like structure the single-mode fiber could be replaced by a mirror placed at the end of the capillary and sealing it off (Fig.\ref{fig:Outlook}b)). In this case the MM fiber has to guide the light to the spectroscopy capillary as well as collect it after the reflection and guide it to a photo-diode to do the signal read out. 
\\
If one wants to increase the flexibility of the sensing area a hollow-core photonic crystal fiber (HC-PCF) could be spliced to the capillary Fig.\ref{fig:Outlook}c) and then either be followed by a SM fiber or a mirror. The transmission properties of the cases b) and c) are of course completely different as for the case described in this publication and have to be examined separately. The working principle of the device however should not change since the filling mechanism of the spectroscopy capillary stays the same. 
\\
\textbf{Acknowledgement}
\\
We acknowledge support from Deutsche Forschungsgemeinschaft (DFG) within the project LO 1657/6-1.

\bibliographystyle{unsrt}
\setlength{\bibsep}{0.0pt}
\bibliography{Bib}

\begin{thebibliography}{10}

\bibitem{Russel}
F.~Benabid, F.~Couny, J.C. Knight, T.A. Birks, and P.~St~J. Russell.
\newblock Compact, stable and efficient all-fibre gas cells using hollow-core
  photonic crystal fibres.
\newblock {\em Nature}, 434:488, 03 2005.

\bibitem{Lukin}
D.~F. Phillips, A.~Fleischhauer, A.~Mair, R.~L. Walsworth, and M.~D. Lukin.
\newblock Storage of light in atomic vapor.
\newblock {\em Phys. Rev. Lett.}, 86:783--786, Jan 2001.

\bibitem{Budker.2007}
Dmitry Budker and Michael Romalis.
\newblock Optical magnetometry.
\newblock {\em Nature Physics}, 3(4):227--234, 2007.

\bibitem{Hofferberth}
M.~Bajcsy, S.~Hofferberth, T.~Peyronel, V.~Balic, Q.~Liang, A.~S. Zibrov,
  V.~Vuletic, and M.~D. Lukin.
\newblock Laser-cooled atoms inside a hollow-core photonic-crystal fiber.
\newblock {\em Phys. Rev. A}, 83:063830, Jun 2011.

\bibitem{Blatt}
Frank Blatt, Thomas Halfmann, and Thorsten Peters.
\newblock One-dimensional ultracold medium of extreme optical depth.
\newblock {\em Opt. Lett.}, 39(3):446--449, Feb 2014.

\bibitem{Benabid2}
C.~Perrella, P.~S. Light, J.~D. Anstie, F.~Benabid, T.~M. Stace, A.~G. White,
  and A.~N. Luiten.
\newblock High-efficiency cross-phase modulation in a gas-filled waveguide.
\newblock {\em Phys. Rev. A}, 88:013819, Jul 2013.

\bibitem{Wamsley2}
M.~R. Sprague, P.~S. Michelberger, T.~F.~M. Champion, D.~G. England, J.~Nunn,
  X.-M. Jin, W.~S. Kolthammer, A.~Abdolvand, P.~St.~J. Russell, and I.~A.
  Walmsley.
\newblock Broadband single-photon-level memory in a hollow-core photonic
  crystal fibre.
\newblock {\em Nat Photon}, 8:287 --291, 2014.

\bibitem{Gaeta2}
V.~Venkataraman, K.~Saha, and A.~L. Gaeta.
\newblock Phase modulation at the few-photon level for weak-nonlinearity-based
  quantum computing.
\newblock {\em Nat Photon}, 7:138--141, 2013.

\bibitem{Shaffer}
Haoquan Fan, Santosh Kumar, Jonathon Sedlacek, Harald K\"{u}bler, Shaya
  Karimkashi, and James~P Shaffer.
\newblock Atom based rf electric field sensing.
\newblock {\em Journal of Physics B: Atomic, Molecular and Optical Physics},
  48(20):202001, 2015.

\bibitem{Weller:17}
Daniel Weller, Arzu Yilmaz, Harald K\"{u}bler, and Robert L\"{o}w.
\newblock High vacuum compatible fiber feedthrough for hot alkali vapor cells.
\newblock {\em Appl. Opt.}, 56(5):1546--1549, Feb 2017.

\bibitem{Gaeta}
Prathamesh~S. Donvalkar, Sven Ramelow, St\'{e}phane Clemmen, and Alexander~L.
  Gaeta.
\newblock Continuous generation of rubidium vapor in hollow-core photonic
  bandgap fibers.
\newblock {\em Opt. Lett.}, 40(22):5379--5382, Nov 2015.

\bibitem{Wamsley}
Krzysztof~T. Kaczmarek, Dylan~J. Saunders, Michael~R. Sprague, W.~Steven
  Kolthammer, Amir Feizpour, Patrick~M. Ledingham, Benjamin Brecht, Eilon Poem,
  Ian~A. Walmsley, and Joshua Nunn.
\newblock Ultrahigh and persistent optical depths of cesium in kagom\'{e}-type
  hollow-core photonic crystal fibers.
\newblock {\em Opt. Lett.}, 40(23):5582--5585, Dec 2015.

\end{thebibliography}

\end{document}